\documentclass[aps,pra,twocolumn,groupedaddress,pdfLaTeX]{revtex4-2}

\usepackage{times}
\usepackage{amsmath}
\usepackage{bm}
\usepackage{graphicx}
\usepackage{hyperref}
\hypersetup{colorlinks=true,citecolor=magenta,linkcolor=magenta,urlcolor=magenta}

\begin{document}


\title{Scaling of errors in digitized counterdiabatic driving}


\author{Takuya Hatomura}
\email[]{takuya.hatomura.ub@hco.ntt.co.jp}
\affiliation{NTT Basic Research Laboratories \& NTT Research Center for Theoretical Quantum Physics, NTT Corporation, Kanagawa 243-0198, Japan}


\date{\today}

\begin{abstract}
We study errors caused by digitization of shortcuts to adiabaticity by counterdiabatic driving. 
We find possibility of error scaling $\mathcal{O}(M^{-2})$ with the number of time slices $M$, whereas worse error scaling $\mathcal{O}(M^{-1})$ is predicted in the conventional theory of the first-order Suzuki-Trotter decomposition. 
We point out this possibility by considering a state-dependent error bound and confirm emergence of this error scaling $\mathcal{O}(M^{-2})$ by numerical simulation. 
Moreover, we numerically show that intermediate error scaling can be observed in digitization of approximate counterdiabatic driving. 
These results reveal usefulness of digitized counterdiabatic driving from the viewpoints of both cost and performance. 
\end{abstract}


\maketitle


%
%
\section{Introduction}

Adiabatic control is a basic approach for tailoring quantum states in desired ways. 
It enables us to track instantaneous energy eigenstates of given systems~\cite{Kato1950}. 
Many quantum algorithms based on adiabatic control have been proposed. 
An advantage of adiabatic control is robustness against errors in control schedules. 
Indeed, we can obtain similar results as long as the adiabatic condition is satisfied (see, Ref.~\cite{Albash2018} and references therein). 
A disadvantage of it is long operation time required to satisfy the adiabatic condition. 
It makes adiabatic algorithms inefficient and exposes them to decoherence.

Speedup of adiabatic control is an important subject. 
Various methods have been proposed for it, e.g., counterdiabatic driving~\cite{Demirplak2003,Berry2009}, invariant-based inverse engineering~\cite{Chen2010}, fast-forward scaling~\cite{Masuda2008,Masuda2010}, quantum adiabatic brachistochrone~\cite{Rezakhani2009}, etc. 
In particular, counterdiabatic driving and invariant-based inverse engineering are collectively referred as shortcuts to adiabaticity and have been extensively studied~\cite{Guery-Odelin2019}.

In counterdiabatic driving, diabatic changes are canceled out by inducing additional driving and fast-forwarded adiabatic time evolution is realized~\cite{Demirplak2003,Berry2009}. 
Although counterdiabatic driving is a promising method, it requires time-dependent control of various non-local and many-body interactions at the same time, and thus practical realization of counterdiabatic driving is still challenging to be applied to adiabatic algorithms.

Recently, digitization of counterdiabatic driving was considered and implemented on quantum devices~\cite{Hegade2021,Chandarana2022,Hegade2022}. 
In digitized counterdiabatic driving, we can independently induce original driving and additional driving. 
Moreover, time dependence of these driving is translated into time duration of constant-strength driving. 
Therefore, digitized counterdiabatic driving is expected as a potential candidate for effectively realizing fast adiabatic algorithms.

Quantifying cost and performance of digitized counterdiabatic driving is an important task for evaluating its practical potential. 
By using a bound for similarity of two quantum dynamics~\cite{Suzuki2020}, performance of approximate shortcuts to adiabaticity can be evaluated with limited knowledge of given systems~\cite{Hatomura2021,Hatomura2022b}. 
However, this evaluation approach cannot directly be applied to digitized counterdiabatic driving. 
The reason being that the bound depends on difference between two Hamiltonians of given dynamics and it does not make sense when difference is large. 
In digital quantum simulation~\cite{Huyghebaert1990,Lloyd1996}, we effectively realize a given Hamiltonian by combination of quantum gate operations, and thus at each time Hamiltonians are completely different. 
Recently, the present author proposed a discrete version of this bound and it can be used for evaluating performance of digital quantum simulation~\cite{Hatomura2022}.

In this paper, we study scaling of errors in digitized counterdiabatic driving against the number of time slices $M$. 
By applying a state-dependent bound to an overlap measure between counterdiabatic driving and digitized counterdiabatic driving, we find that the second-order error terms $\mathcal{O}(M^{-2})$ vanish, which are consistent with the dominant error terms in the first-order Suzuki-Trotter decomposition~\cite{Suzuki1976}. 
As a result, the third-order error terms $\mathcal{O}(M^{-3})$ become dominant terms. 
Because these errors accumulate for $M$ time slices, we conclude that scaling of errors in digitized counterdiabatic driving is $\mathcal{O}(M^{-2})$. 
We numerically confirm this error scaling and also study scaling of errors in digitized approximate-counterdiabatic driving.

%
%
\section{Background}

In this section, we give an overview of previous works. 
We summarize digitized dynamics in Sec.~\ref{Sec.dd}, a method for evaluating its errors in Sec.~\ref{Sec.errors}, and digitized counterdiabatic driving in Sec.~\ref{Sec.dcd}.

%
%
\subsection{\label{Sec.dd}Digitized dynamics}

We consider dynamics $|\Psi(t)\rangle=\hat{U}(t,0)|\Psi(0)\rangle$ under a Hamiltonian $\hat{\mathcal{H}}(t)$, where $\hat{U}(t,0)$ is a time-evolution operator 
\begin{equation}
\hat{U}(t,0)=\mathcal{T}\exp\left(-\frac{i}{\hbar}\int_0^tdt^\prime\hat{\mathcal{H}}(t^\prime)\right).
\label{Eq.tevo}
\end{equation}
We assume that the Hamiltonian consists of simulatable parts as
\begin{equation}
\hat{\mathcal{H}}(t)=\sum_k\hat{H}_k(t), 
\end{equation}
where each $\hat{H}_k(t)$ can be simulated on quantum devices.

In digital quantum simulation~\cite{Huyghebaert1990,Lloyd1996}, we approximate the time-evolution operator (\ref{Eq.tevo}) as
\begin{equation}
\hat{U}(T,0)\approx\prod_{n=M}^1\hat{U}_d\bm{(}nT/M,(n-1)T/M\bm{)}, 
\label{Eq.tevo.discretization}
\end{equation}
where $T$ is the final time, $M$ is the number of time slices,
\begin{equation}
\hat{U}_d\bm{(}nT/M,(n-1)T/M\bm{)}=\prod_k\hat{U}_k\bm{(}nT/M,(n-1)T/M\bm{)},
\label{Eq.tevo.Trotterization}
\end{equation}
and
\begin{equation}
\hat{U}_k\bm{(}nT/M,(n-1)T/M\bm{)}=\exp\left(-\frac{i}{\hbar}\frac{T}{M}\hat{H}_k(nT/M)\right). 
\label{Eq.tevo.simulatable}
\end{equation}
Here, the first-order Suzuki-Trotter decomposition is adopted and its errors scale as $\mathcal{O}(M^{-1})$. 
Precisely, there are $M$ time slices and each slice causes $\mathcal{O}(M^{-2})$ errors. 
Note that approximation $\hat{\mathcal{H}}(nT/M)\approx\hat{\mathcal{H}}\bm{(}(n-1)T/M\bm{)}$ is also assumed, but its errors also scale as $\mathcal{O}(M^{-2})$ for each slice. 
Then, digitized dynamics
\begin{equation}
|\Psi_d(mT/M)\rangle=\prod_{n=m}^1\hat{U}_d\bm{(}nT/M,(n-1)T/M\bm{)}|\Psi_d(0)\rangle, 
\end{equation}
is realized, where $m=1,2,\dots,M$ and $|\Psi_d(0)\rangle$ is the initial state. 
We assume the identical initial states, $|\Psi_d(0)\rangle=|\Psi(0)\rangle$.

%
%
\subsection{\label{Sec.errors}Errors in digitized dynamics}

According to Ref.~\cite{Hatomura2022}, we can obtain inequality
\begin{equation}
|\langle\Psi(T)|\Psi_d(T)\rangle|\ge\cos\left(\sum_{n=1}^M\mathcal{L}_n\right),\quad\text{for }\sum_{n=1}^M\mathcal{L}_n\le\frac{\pi}{2},
\label{Eq.bound}
\end{equation}
where
\begin{equation}
\begin{aligned}
\mathcal{L}_n=\arccos&|\langle\Psi(nT/M)|\hat{U}_d\bm{(}nT/M,(n-1)T/M\bm{)} \\
&\times[\hat{U}\bm{(}nT/M,(n-1)T/M\bm{)}]^\dag|\Psi(nT/M)\rangle|. 
\end{aligned}
\label{Eq.Ln}
\end{equation}
Note that the roles of reference dynamics and digitized dynamics in Eq.~(\ref{Eq.Ln}) can be exchanged. 
By applying the Taylor expansion to Eq.~(\ref{Eq.Ln}), we find
\begin{equation}
\mathcal{L}_n\approx\frac{T^2}{2\hbar^2M^2}|\langle\Psi(nT/M)|\hat{A}(nT/M)|\Psi(nT/M)\rangle|,
\label{Eq.Ln.approx}
\end{equation}
where
\begin{equation}
\hat{A}(nT/M)=\sum_{\substack{k,l \\ (k\neq l)}}[\hat{H}_k(nT/M),\hat{H}_l(nT/M)].
\end{equation}
Since there are $M$ slices of $\mathcal{L}_n$, this bound gives the identical error scaling $\mathcal{O}(M^{-1})$, but it depend on a given state and is tighter than conventional analyses~\cite{Huyghebaert1990,Lloyd1996}.

%
%
\subsection{\label{Sec.dcd}Digitized counterdiabatic driving}

We consider a reference Hamiltonian
\begin{equation}
\hat{\mathcal{H}}_\mathrm{ref}(t)=\sum_nE_n(t)|n(t)\rangle\langle n(t)|, 
\label{Eq.refham}
\end{equation}
where $E_n(t)$ is an eigen-energy and $|n(t)\rangle$ is its energy eigenstate. 
According to the theory of counterdiabatic driving~\cite{Demirplak2003,Berry2009}, we can realize adiabatic time evolution of the reference Hamiltonian (\ref{Eq.refham}) by inducing the counterdiabatic Hamiltonian
\begin{equation}
\hat{\mathcal{H}}_\mathrm{cd}(t)=i\hbar\sum_n\bm{(}1-|n(t)\rangle\langle n(t)|\bm{)}|\partial n(t)\rangle\langle n(t)|,
\label{Eq.cdham}
\end{equation}
where $\partial=\partial/\partial t$. 
The time-evolution operator of counterdiabatic driving is given by
\begin{equation}
\hat{U}_\mathrm{ad}(t,0)=\mathcal{T}\exp\left(-\frac{i}{\hbar}\int_0^tdt^\prime\hat{\mathcal{H}}_\mathrm{ad}(t^\prime)\right),
\label{Eq.adtevo}
\end{equation}
where
\begin{equation}
\hat{\mathcal{H}}_\mathrm{ad}(t)=\hat{\mathcal{H}}_\mathrm{ref}(t)+\hat{\mathcal{H}}_\mathrm{cd}(t). 
\label{Eq.adham}
\end{equation}
The adiabatic state is given by $|\Psi_\mathrm{ad}(t)\rangle=\hat{U}_\mathrm{ad}(t,0)|\Psi_\mathrm{ad}(0)\rangle$.

In digitized counterdiabatic driving~\cite{Hegade2021,Chandarana2022,Hegade2022}, we divide the time-evolution operator~(\ref{Eq.adtevo}) of the total Hamiltonian (\ref{Eq.adham}) into that of the reference Hamiltonian (\ref{Eq.refham}) and that of the counterdiabatic Hamiltonian (\ref{Eq.cdham}) as Eqs.~(\ref{Eq.tevo.discretization}), (\ref{Eq.tevo.Trotterization}), and (\ref{Eq.tevo.simulatable}). 
Here, the resulting dynamics is denoted as $|\Psi_{\mathrm{ad},d}(mT/M)\rangle$ for $m=1,2,\dots,M$.

%
%
\section{Results}

In this section, we show our results. 
We discuss possibility of error scaling $\mathcal{O}(M^{-2})$ in Sec.~\ref{Sec.errors.dcd} and show emergence of it by numerical simulation in Sec.~\ref{Sec.numerical}. 
Moreover, we also discuss digitized approximate-counterdiabatic driving in Sec.~\ref{Sec.numerical}.

%
%
\subsection{\label{Sec.errors.dcd}Possibility of error scaling $\mathcal{O}(M^{-2})$}

Many quantum algorithms only use a single energy eigenstate of the reference Hamiltonian (\ref{Eq.refham}), and thus we assume that the initial state is given by $|\Psi_\mathrm{ad}(0)\rangle=|g(0)\rangle$, where $|g(0)\rangle$ is one of the energy eigenstates of the reference Hamiltonian (\ref{Eq.refham}). 
Then, dynamics by counterdiabatic driving tracks the single energy eigenstate as
\begin{equation}
|\Psi_\mathrm{ad}(t)\rangle=e^{-\frac{i}{\hbar}\int_0^tdt^\prime E_g(t^\prime)}e^{i\gamma_g(t)}|g(t)\rangle,
\label{Eq.adstate}
\end{equation}
where $\gamma_g(t)$ is the Berry phase. 
Remarkably, we find that the right-hand side of Eq.~(\ref{Eq.Ln.approx}) vanishes when we substitute Eqs.~(\ref{Eq.refham}), (\ref{Eq.cdham}), and (\ref{Eq.adstate}) for it. 
It means that errors in digitized counterdiabatic driving may not scale as $\mathcal{O}(M^{-1})$ even if we adopt the simplest decomposition, i.e., the first-order Suzuki-Trotter decomposition.

To find dominant errors in digitized counterdiabatic driving, we study higher-order terms in the Taylor expansion of Eq.~(\ref{Eq.Ln}). 
As a result, we find
\begin{equation}
\mathcal{L}_n\approx\frac{T^3}{6\hbar^3M^3}|\langle g(nT/M)|\hat{B}(nT/M)|g(nT/M)\rangle|,
\label{Eq.Ln.dcd}
\end{equation}
where
\begin{equation}
\hat{B}(nT/M)=\bm{[}\hat{\mathcal{H}}_\mathrm{cd}(nT/M),[\hat{\mathcal{H}}_\mathrm{cd}(nT/M),\hat{\mathcal{H}}_\mathrm{ref}(nT/M)]\bm{]}.
\end{equation}
By using the expressions (\ref{Eq.refham}) and (\ref{Eq.cdham}), we can also rewrite it as
\begin{equation}
\mathcal{L}_n\approx\frac{T^3}{3\hbar M^3}\left|\sum_{\substack{k \\ (k\neq g)}}\frac{\langle g|(\partial\hat{\mathcal{H}}_\mathrm{ref})|k\rangle\langle k|(\partial\hat{\mathcal{H}}_\mathrm{ref})|g\rangle}{E_k-E_g}\right|, 
\label{Eq.Ln.dcd2}
\end{equation}
where arguments $(nT/M)$ are omitted for simplicity. 
This term gives $\mathcal{O}(M^{-3})$ error scaling and we have $M$ slices of $\mathcal{L}_n$, and thus we expect that errors in digitized counterdiabatic driving scale as $\mathcal{O}(M^{-2})$.

We remark that emergence of the error scaling $\mathcal{O}(M^{-2})$ is just possibility. 
In discretization of counterdiabatic driving, approximation $\hat{\mathcal{H}}_\mathrm{ad}(nT/M)\approx\hat{\mathcal{H}}_\mathrm{ad}\bm{(}(n-1)T/M\bm{)}$ is assumed and it causes $\mathcal{O}(M^{-2})$ errors for each slice. 
Moreover, the bound (\ref{Eq.bound}) with Eq.~(\ref{Eq.Ln.dcd}) [(\ref{Eq.Ln.dcd2})] just gives the worst-case performance of digitization and it does not guarantee scaling of actual dynamics. 
The error scaling $\mathcal{O}(M^{-2})$ can be obtained when these influences are subdominant.

%
%
\subsection{\label{Sec.numerical}Emergence of error scaling $\mathcal{O}(M^{-2})$}

\begin{figure}
\includegraphics[width=8cm]{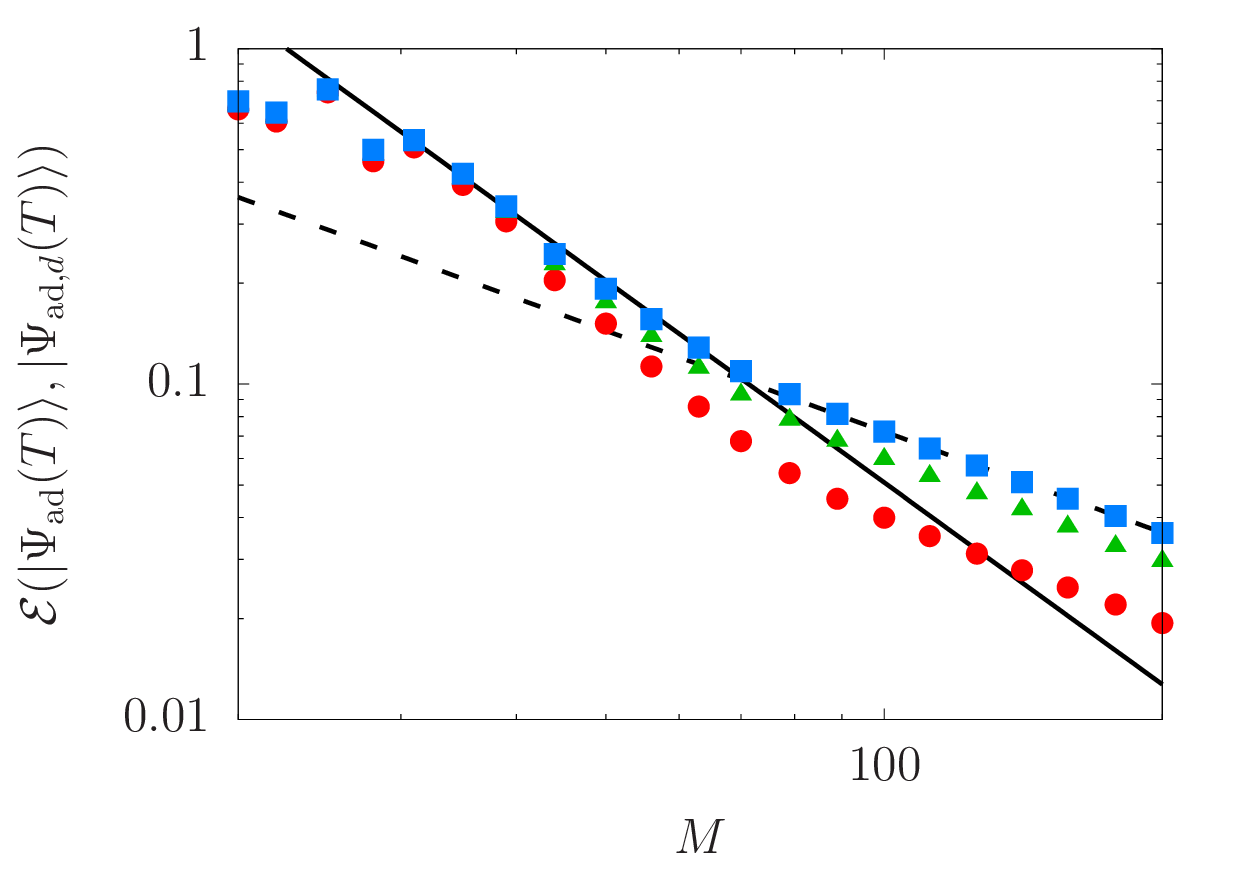}
\includegraphics[width=8cm]{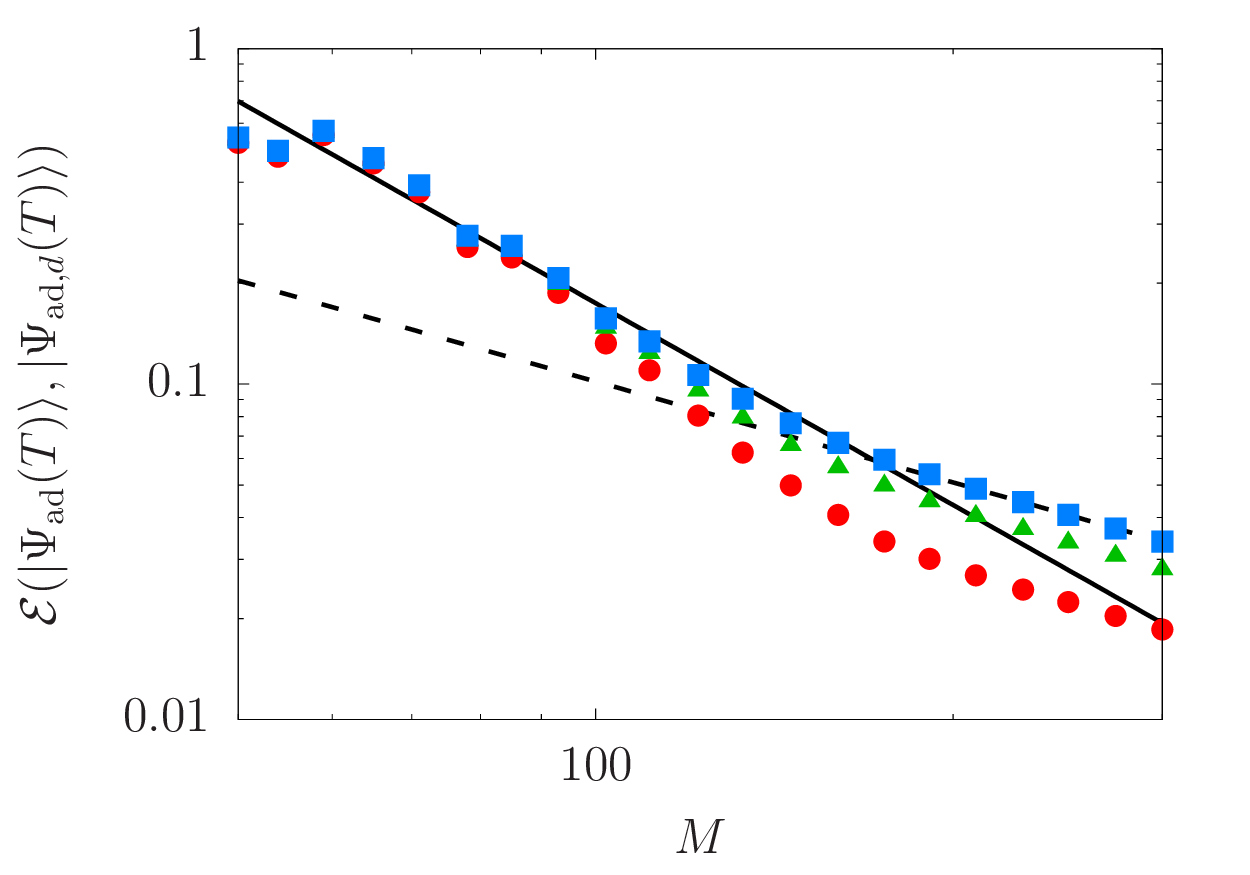}
\caption{\label{Fig.infidelity} Infidelity of digitized counterdiabatic driving to counterdiabatic driving, $\mathcal{E}\bm{(}|\Psi_\mathrm{ad}(T)\rangle,|\Psi_{\mathrm{ad},d}(T)\rangle\bm{)}$, against the number of time slices $M$. Plotted symbols represent (red circles) $T=1$, (green triangles) $T=5$, and (blue squares) $T=10$, respectively. Error scalings $\mathcal{O}(M^{-2})$ and $\mathcal{O}(M^{-1})$ are indicated by the black solid line and the dashed one, respectively. The plotted ranges of the $x$-axis are (top) $[20,200]$ and (bottom) $[50,300]$, respectively. Here, (top) $L=50$ and (bottom) $L=100$. }
\end{figure}

\begin{figure}
\includegraphics[width=8cm]{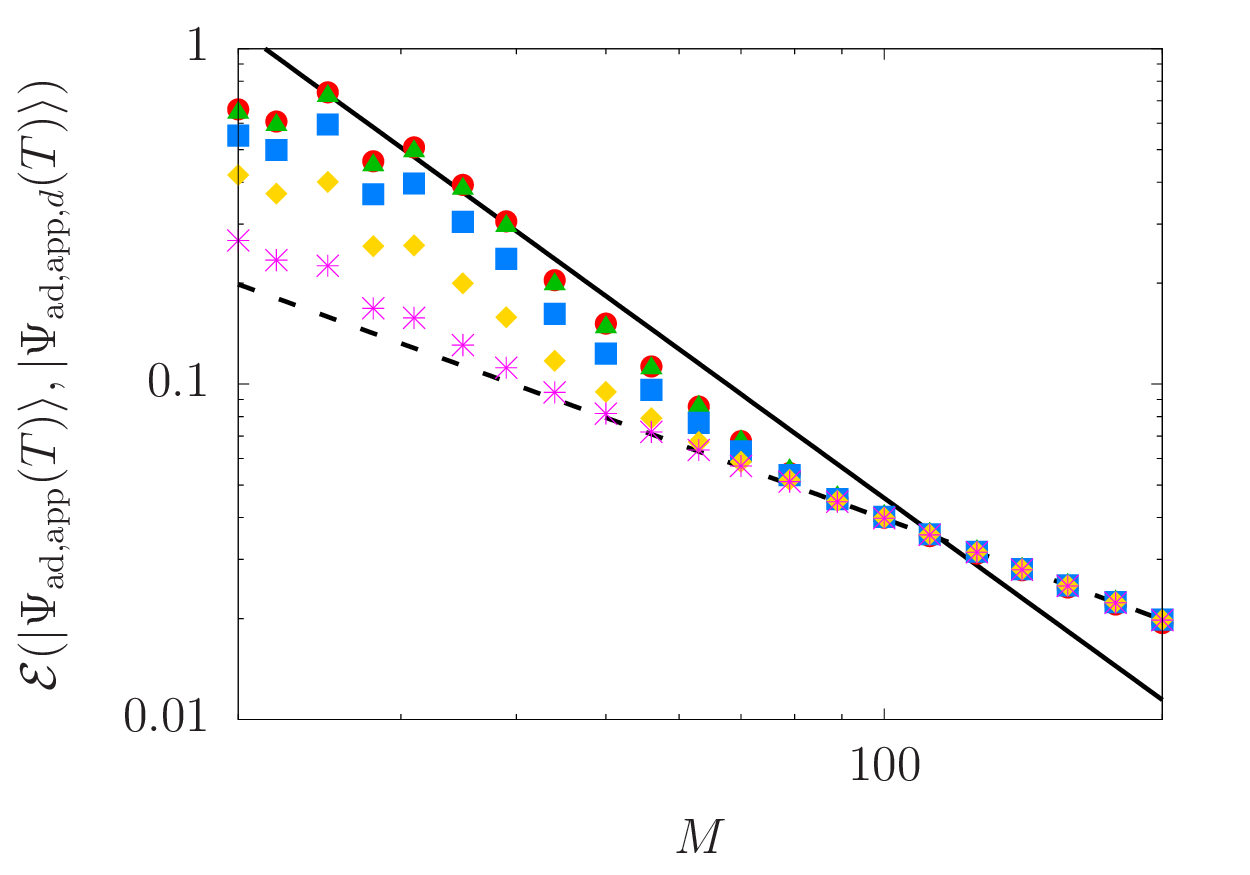}
\caption{\label{Fig.trunc} Infidelity of digitized approximate-counterdiabatic driving to approximate counterdiabatic driving, $\mathcal{E}\bm{(}|\Psi_\mathrm{ad,app}(T)\rangle,|\Psi_{\mathrm{ad,app},d}(T)\rangle\bm{)}$, against the number of time slices $M$. Plotted symbols represent the number of truncated interactions (red circles) 0, (green triangles) 10, (blue squares) 20, (yellow diamonds) 30, and (magenta asterisks) 40, respectively. Error scalings $\mathcal{O}(M^{-2})$ and $\mathcal{O}(M^{-1})$ are indicated by the black solid line and the dashed one, respectively. The plotted range of the $x$-axis is $[20,200]$. Here, $T=1$ and $L=50$. }
\end{figure}

\begin{figure}
\includegraphics[width=8cm]{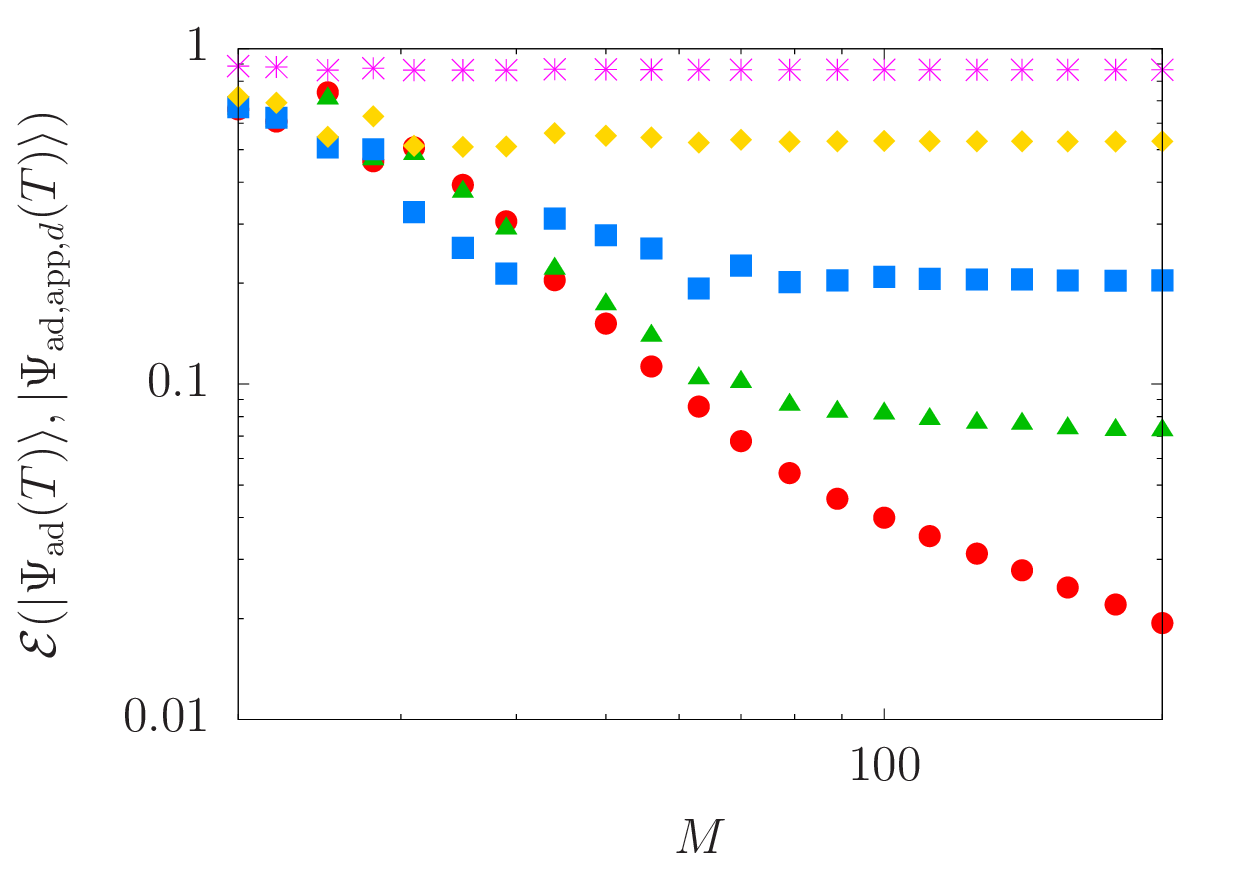}
\caption{\label{Fig.GSinfid} Infidelity of digitized approximate-counterdiabatic driving to the adiabatic state, $\mathcal{E}\bm{(}|\Psi_\mathrm{ad}(T)\rangle,|\Psi_{\mathrm{ad,app},d}(T)\rangle\bm{)}$, against the number of time slices $M$. Plotted symbols represent the number of truncated interactions (red circles) 0, (green triangles) 10, (blue squares) 20, (yellow diamonds) 30, and (magenta asterisks) 40, respectively. The plotted range of the $x$-axis is $[20,200]$. Here, $T=1$ and $L=50$. }
\end{figure}

For numerical simulation, we consider quantum annealing in the transverse Ising chain~\cite{Kadowaki1998}. 
Its reference Hamiltonian is given by
\begin{equation}
\hat{\mathcal{H}}_\mathrm{ref}(t)=\lambda(t)\hat{\mathcal{H}}_P+\bm{(}1-\lambda(t)\bm{)}\hat{\mathcal{H}}_V,
\label{Eq.QAham}
\end{equation}
where $\hat{\mathcal{H}}_P$ is a problem Hamiltonian, which is the one-dimensional Ising model
\begin{equation}
\hat{\mathcal{H}}_P=-J\sum_{i=1}^L\hat{Z}_i\hat{Z}_{i+1}, 
\label{Eq.proham}
\end{equation}
with the periodic boundary condition $\hat{Z}_{L+1}=\hat{Z}_1$, and $\hat{\mathcal{H}}_V(t)$ is a driver Hamiltonian, which is a transverse-field Hamiltonian
\begin{equation}
\hat{\mathcal{H}}_V=-h^x\sum_{i=1}^L\hat{X}_i. 
\label{Eq.transham}
\end{equation}
Here, $\{\hat{X}_i,\hat{Y}_i,\hat{Z}_i\}_{i=1,2,\dots,L}$ denotes the Pauli matrices of $L$ qubits, $J$ is the strength of interaction, and $h^x$ is that of the transverse field. 
The trivial ground state of the driver Hamiltonian and the ground state of the problem Hamiltonian are interpolated by a time-dependent parameter $\lambda(t)$, and thus it must satisfy $\lambda(0)=0$ and $\lambda(T)=1$. 
Here and hereafter, we set $J=h^x=1$ and $\hbar=1$ to adopt the conventional notation of quantum annealing, and then timescale becomes dimensionless.

The counterdiabatic Hamiltonian for the reference Hamiltonian (\ref{Eq.QAham}) with Eqs.~(\ref{Eq.proham}) and (\ref{Eq.transham}) is given by
\begin{equation}
\hat{\mathcal{H}}_\mathrm{cd}(t)=\sum_{k=0}^K\alpha_k(t)\sum_{i=1}^L(\hat{Y}_i\hat{X}_i^{(k)}\hat{Z}_{i+k+1}+\hat{Z}_i\hat{X}_i^{(k)}\hat{Y}_{i+k+1}),
\label{Eq.isingcd}
\end{equation}
where
\begin{equation}
\hat{X}_i^{(k)}=\prod_{j=1}^k\hat{X}_{i+j},\quad\text{for }k=1,2,\dots,L-2,
\end{equation}
and $\hat{X}_i^{(0)}=1$~\cite{delCampo2012,Damski2014}. 
In this paper, we numerically determine its time-dependent coefficient $\alpha_k(t)$ by using the algebraic approach~\cite{Hatomura2021}. 
Its explicit expression can be found in Refs.~\cite{delCampo2012,Damski2014}, whereas modification is necessary because they adopt time-independent interaction. 
Here, $K$ is an integer, $K\le L-2$, which is later used to discuss digitized approximate-counterdiabatic driving by truncating higher-order terms. 
In exact counterdiabatic driving, $K=L-2$.

In numerical simulation, we adopt a schedule $\lambda(t)=(1/2)[1-\cos(\pi t/T)]$, with which the counterdiabatic Hamiltonian (\ref{Eq.cdham}) vanishes at the initial and final time, and evaluate infidelity
\begin{equation}
\mathcal{E}\bm{(}|\Psi_1(T)\rangle,|\Psi_2(T)\rangle\bm{)}=\sqrt{1-|\langle\Psi_1(T)|\Psi_2(T)\rangle|^2}, 
\end{equation}
where $|\Psi_1(T)\rangle$ and $|\Psi_2(T)\rangle$ is given two states.

First, we evaluate the infidelity of digitized counterdiabatic driving to counterdiabatic driving, $\mathcal{E}\bm{(}|\Psi_\mathrm{ad}(T)\rangle,|\Psi_{\mathrm{ad},d}(T)\rangle\bm{)}$. 
We plot it against the number of time slices $M$ in Fig.~\ref{Fig.infidelity}. 
When the number of time slices $M$ is not very large, we clearly find reduction of the infidelity scaling as $\mathcal{O}(M^{-2})$. 
We also find that scaling of errors becomes $\mathcal{O}(M^{-1})$ for large $M$. 
This result can simply be understood as that dominant errors scaling as $\mathcal{O}(M^{-2})$ quickly decrease, but subdominant errors scaling as $\mathcal{O}(M^{-1})$ do not. 
As a result, dominance is exchanged at a certain point. 
One of the candidates for subdominant errors is approximation $\hat{\mathcal{H}}_\mathrm{ad}(nT/M)\approx\hat{\mathcal{H}}_\mathrm{ad}\bm{(}(n-1)T/M\bm{)}$. 
We show evidence of this possibility below.

Next, we consider digitized approximate-counterdiabatic driving. 
Here we assume $K\le L-2$ in Eq.~(\ref{Eq.isingcd}). 
Dynamics by approximate counterdiabatic driving is denoted as $|\Psi_\mathrm{ad,app}(T)\rangle$ and its digitized version is denoted as $|\Psi_{\mathrm{ad,app},d}(T)\rangle$. 
We evaluate the infidelity of digitized approximate-counterdiabatic driving to approximate counterdiabatic driving, $\mathcal{E}\bm{(}|\Psi_\mathrm{ad,app}(T)\rangle,|\Psi_{\mathrm{ad,app},d}(T)\rangle\bm{)}$. 
We plot it against $M$ in Fig.~\ref{Fig.trunc}. 
We find that scaling of errors gradually tends to $\mathcal{O}(M^{-1})$ when we increase the number of truncated interactions. 
We also find that it converges on the line which is found in Fig.~\ref{Fig.infidelity} for large $M$. 
That is, it is evident that subdominant errors in Fig.~\ref{Fig.infidelity} comes from approximation $\hat{\mathcal{H}}_\mathrm{ad}(nT/M)\approx\hat{\mathcal{H}}_\mathrm{ad}\bm{(}(n-1)T/M\bm{)}$ in discretization, or more specifically, that of the reference Hamiltonian $\hat{\mathcal{H}}_\mathrm{ref}(nT/M)\approx\hat{\mathcal{H}}_\mathrm{ref}\bm{(}(n-1)T/M\bm{)}$. 
Note that the infidelity becomes small when we increase the number of truncated interactions, but it should come from the fact that the number of involved interactions becomes small and it does not mean better performance. 
We remark this below.

Finally, we consider the infidelity of digitized approximate-counterdiabatic driving to the adiabatic state (exact counterdiabatic driving), $\mathcal{E}\bm{(}|\Psi_\mathrm{ad}(T)\rangle,|\Psi_{\mathrm{ad,app},d}(T)\rangle\bm{)}$. 
We plot it against $M$ in Fig.~\ref{Fig.GSinfid}. 
We find that the infidelity to the adiabatic state decreases in the similar way to the exact one for small $M$, i.e., better than $\mathcal{O}(M^{-1})$, even if some higher-order terms are truncated although it saturates at certain points because of imperfection of the counterdiabatic Hamiltonian.

%
%
\section{Summary}

In this paper, we studied scaling of errors in digitized counterdiabatic driving. 
We found that digitization errors of counterdiabatic driving scale as $\mathcal{O}(M^{-2})$, whereas the conventional prediction is $\mathcal{O}(M^{-1})$. 
We also showed that digitization of approximate counterdiabatic driving causes intermediate errors between $\mathcal{O}(M^{-2})$ and $\mathcal{O}(M^{-1})$. 
These findings enhance usefulness of digitized counterdiabatic driving from the both viewpoints of cost and performance.

When we implement digitized counterdiabatic driving on real quantum devices, we may have to further decompose the time-evolution operators of the reference Hamiltonian and the counterdiabatic Hamiltonian. 
Not to smear beneficial error scaling $\mathcal{O}(M^{-2})$, we should adopt the second-order Suzuki-Trotter decomposition~\cite{Suzuki1985} or other techniques. 
Notably, we can also adopt efficient product formulae for decomposing the time-evolution operator of the (approximate) counterdiabatic Hamiltonian instead of the second-order Suzuki-Trotter decomposition~\cite{Childs2013,Chen2022}.

\bibliography{Trotter_adcd_bib.bib}

\end{document}